\documentstyle[12pt]{article}

\begin{document}

\title{\Large {\bf Anticommutators and propagators of Moyal 
        star-products for Dirac field on noncommutative spacetime } }
  
\author{{{\large Zheng Ze Ma}} \thanks{Electronic address: 
           z.z.ma@seu.edu.cn} 
  \\  \\ {\normalsize {\sl Department of Physics, Southeast University, 
         Nanjing, 210096, P. R. China } }}

\date{}

\maketitle

\begin{abstract}

\indent

  We study the Moyal anticommutators and their expectation values 
between vacuum states and non-vacuum states for Dirac fields on 
noncommutative spacetime. Then we construct the propagators of 
Moyal star-products for Dirac fields on noncommutative spacetime. We 
find that the propagators of Moyal star-products for Dirac fields are  
equal to the propagators of Dirac fields on ordinary commutative 
spacetime. 

\end{abstract}

~~~ PACS numbers: 11.10.Nx, 03.70.+k

\vskip 0.5cm

\indent

  Spacetime may have discrete and noncommutative structures under a 
very small microscopic scale. The concept of spacetime noncommutativity 
was first proposed by Snyder many years ago [1]. The purely mathematical 
development on noncommutative geometry was carried out by Connes [2]. 
To combine Heisenberg's uncertainty principle and Einstein's 
gravitational equations, Doplicher {\sl et al}. proposed the uncertainty 
relations for the measurement of spacetime coordinates [3]. In recent 
years, spacetime noncommutativity was discovered again in superstring 
theories [4]. It has resulted a lot of researches on noncommutative 
field theories [5,6].

  In noncommutative spacetime, we can regard the Moyal star-product as 
the basic product operation. Therefore we need to study the commutators 
and anticommutators and propagators of Moyal star-products for quantum 
fields on noncommutative spacetime. In Ref. [7] we studied the 
commutators and propagators of Moyal star-products for noncommutative 
scalar field theory. In this paper, we will study the anticommutators and 
propagators of Moyal star-products for noncommutative Dirac field.

  In noncommutative spacetime, spacetime coordinates satisfy the 
commutation relation
$$
  [x^{\mu},x^{\nu}]=i\theta^{\mu\nu} ~,
  \eqno{(1)}  $$
where $\theta^{\mu\nu}$ is a constant real antisymmetric matrix that 
parameterizes the noncommutativity of the spacetime. There is a map 
between field theories on ordinary spacetime and on noncommutative 
spacetime. For field theories on noncommutative spacetime, they can be 
obtained through introducing the Moyal star-product, i.e., all of the 
products between field functions are replaced by the Moyal star-products. 
The Moyal star-product of two functions $f(x)$ and $g(x)$ is defined to be 
\begin{eqnarray*}
  f(x)\star g(x) & = & e^{\frac{i}{2}\theta^{\mu\nu}\frac{\partial}{\partial 
                 \alpha^\mu}\frac{\partial}{\partial\beta^{\nu}}}f(x+\alpha)
                 g(x+\beta)\vert_{\alpha=\beta=0} \\
                 & = & f(x)g(x)+\sum\limits^{\infty}_{n=1}
                 \left(\frac{i}{2}\right)^{n}
                 \frac{1}{n!}\theta^{\mu_{1}\nu_{1}}
                 \cdots\theta^{\mu_{n}\nu_{n}}
                 \partial_{\mu_{1}}\cdots\partial_{\mu_{n}}f(x) 
                 \partial_{\nu_{1}}\cdots\partial_{\nu_{n}}g(x) ~.
\end{eqnarray*}
$$ 
  \eqno{(2)}  $$
The Moyal star-product of two functions of Eq. (2) is defined at the 
same spacetime point. We can generalize Eq. (2) to two functions on 
different spacetime points [6]:
\begin{eqnarray*}
  f(x_{1})\star g(x_{2}) & = & e^{\frac{i}{2}
          \theta^{\mu\nu}\frac{\partial}{\partial 
          \alpha^\mu}\frac{\partial}{\partial\beta^{\nu}}}f(x_{1}+\alpha)
                 g(x_{2}+\beta)\vert_{\alpha=\beta=0} \\
                 & = & f(x_{1})g(x_{2})+\sum\limits^{\infty}_{n=1}
                 \left(\frac{i}{2}\right)^{n}
                 \frac{1}{n!}\theta^{\mu_{1}\nu_{1}}
                 \cdots\theta^{\mu_{n}\nu_{n}}
                 \partial_{\mu_{1}}\cdots\partial_{\mu_{n}}f(x_{1}) 
                 \partial_{\nu_{1}}\cdots\partial_{\nu_{n}}g(x_{2}) ~. 
\end{eqnarray*}
$$ 
  \eqno{(3)}  $$
Equation (3) can be established through generalize the commutation 
relation of spacetime coordinates at the same point to two different 
points: 
$$
  [x_{1}^{\mu},x_{2}^{\nu}]=i\theta^{\mu\nu} ~. 
  \eqno{(4)}  $$
We can also expect to search for a grounds of argument for Eq. (4) 
from superstring theories.

  To consider the free Dirac field on noncommutative spacetime, its 
Lagrangian is given by 
$$
  {\cal L}=\overline{\psi}\star i\gamma^{\mu}\partial_{\mu}\psi-
            m\overline{\psi}\star\psi ~.
  \eqno{(5)}  $$ 
The Fourier expansions for the free Dirac fields are given by 
$$
  \psi({\bf x},t)=\int\frac{d^{3}p}{(2\pi)^{3/2}}\sqrt
      {\frac{m}{E_{p}}}\sum\limits_{s=1,2}[b(p,s)u(p,s)
      e^{-ip\cdot x}+d^{\dagger}(p,s)v(p,s)e^{ip\cdot x}] ~,      $$
$$
  \overline{\psi}({\bf x},t)=\int\frac{d^{3}p}{(2\pi)^{3/2}}\sqrt
      {\frac{m}{E_{p}}}\sum\limits_{s=1,2}[b^{\dagger}(p,s)
       \overline{u}(p,s)e^{ip\cdot x}+d(p,s)\overline{v}(p,s)
       e^{-ip\cdot x}] ~,      
  \eqno{(6)}  $$ 
where $E_{p}=p_{0}=+\sqrt{\vert{\bf p}\vert^{2}+m^{2}}$. In Eq. (6), 
the spacetime coordinates are treated as noncommutative. They satisfy 
the commutation relations (1) and (4). The commutation relations for 
the creation and annihilation operators are still the same as in the 
commutative spacetime: 
$$
  \{b(p,s),b^{\dagger}(p^{\prime},s^{\prime})\}=\delta_{ss^{\prime}}
           \delta^{3}({\bf p}-{\bf p}^{\prime}) ~,             $$     
$$
  \{d(p,s),d^{\dagger}(p^{\prime},s^{\prime})\}=\delta_{ss^{\prime}}
           \delta^{3}({\bf p}-{\bf p}^{\prime}) ~,             $$    
$$
  \{b(p,s),b(p^{\prime},s^{\prime})\}=
           \{d(p,s),d(p^{\prime},s^{\prime})\}=0 ~,            $$
$$
  \{b^{\dagger}(p,s),b^{\dagger}(p^{\prime},s^{\prime})\}=
       \{d^{\dagger}(p,s),d^{\dagger}(p^{\prime},s^{\prime})\}=0 ~,  $$
$$
  \{b(p,s),d(p^{\prime},s^{\prime})\}=
           \{b(p,s),d^{\dagger}(p^{\prime},s^{\prime})\}=0 ~,        $$
$$
  \{d(p,s),b(p^{\prime},s^{\prime})\}=
           \{d(p,s),b^{\dagger}(p^{\prime},s^{\prime})\}=0 ~.        
  \eqno{(7)}  $$ 
The spinors $u(p,s)$ and $v(p,s)$ satify 
$$
  \sum\limits_{s=1,2}u_{\alpha}(p,s)\overline{u}_{\beta}(p,s)
         =\left(\frac{\not{p}+m}{2m}\right)_{\alpha\beta} ~,        $$
$$
  \sum\limits_{s=1,2}v_{\alpha}(p,s)\overline{v}_{\beta}(p,s)
         =\left(\frac{\not{p}-m}{2m}\right)_{\alpha\beta} ~.          
  \eqno{(8)}  $$

  We define the anticommutator of the Moyal star-product for Dirac field 
to be 
$$
  \{\psi_{\alpha}(x),\overline{\psi}_{\beta}(x^{\prime})\}_{\star}
     =\psi_{\alpha}(x)\star\overline{\psi}_{\beta}(x^{\prime})+
      \overline{\psi}_{\beta}(x^{\prime})\star\psi_{\alpha}(x) ~.      
  \eqno{(9)}  $$ 
Equation (9) can be called the Moyal anticommutator for convenience. From 
the Fourier expansion of Eq. (6) for the free Dirac fields, we can 
calculate the Moyal anticommutator of Eq. (9). It is given by 
\begin{eqnarray*}
  \{\psi_{\alpha}(x),\overline{\psi}_{\beta}(x^{\prime})\}_{\star}  
          & = & \int\frac{d^{3}pd^{3}p^{\prime}}
                {(2\pi)^{3}}\frac{m}{\sqrt{E_{p}E_{p^{\prime}}}}
            \Big\{\sum\limits_{s=1,2}[b(p,s)u_{\alpha}(p,s)e^{-ip\cdot x}+
              d^{\dagger}(p,s)v_{\alpha}(p,s)e^{ip\cdot x}],    \\
          & ~ &  ~~~~~~~~~~~~~~~~ 
                 \sum\limits_{s^{\prime}=1,2}[b^{\dagger}
                (p^{\prime},s^{\prime})\overline{u}_{\beta}
                (p^{\prime},s^{\prime})
                 e^{ip^{\prime}\cdot x^{\prime}}+d(p^{\prime},s^{\prime})
                 \overline{v}_{\beta}(p^{\prime},s^{\prime})e^{-ip^{\prime}
                 \cdot x^{\prime}}]\Big\}_\star      \\ 
           & = & \int\frac{d^{3}pd^{3}p^{\prime}}
                {(2\pi)^{3}}\frac{m}{\sqrt{E_{p}E_{p^{\prime}}}}
                \sum\limits_{s,s^{\prime}}\Big[\Big\{b(p,s)u_{\alpha}(p,s)
                 e^{-ip\cdot x},  
                 b^{\dagger}(p^{\prime},s^{\prime})\overline{u}_{\beta}
                (p^{\prime},s^{\prime})
                 e^{ip^{\prime}\cdot x^{\prime}}\Big\}_\star   \\
           & ~ &  ~~~~~~~~~~~~~~~~     
                +\Big\{b(p,s)u_{\alpha}(p,s)
                 e^{-ip\cdot x},d(p^{\prime},s^{\prime})
                 \overline{v}_{\beta}(p^{\prime},s^{\prime})e^{-ip^{\prime}
                 \cdot x^{\prime}}\Big\}_\star       \\   
           & ~ &  ~~~~~~~~~~~~~~~~     
                +\Big\{d^{\dagger}(p,s)v_{\alpha}(p,s)e^{ip\cdot x}, 
                 b^{\dagger}(p^{\prime},s^{\prime})\overline{u}_{\beta}
                (p^{\prime},s^{\prime})
                 e^{ip^{\prime}\cdot x^{\prime}}\Big\}_\star   \\
           & ~ &  ~~~~~~~~~~~~~~~~    
                +\Big\{d^{\dagger}(p,s)v_{\alpha}(p,s)e^{ip\cdot x}, 
                 d(p^{\prime},s^{\prime})
                 \overline{v}_{\beta}(p^{\prime},s^{\prime})e^{-ip^{\prime}
                 \cdot x^{\prime}}\Big\}_\star\Big] ~.    
\end{eqnarray*}      
$$  \eqno{(10)}  $$
In Eq. (10) there exist two kinds of noncommutative structures, field 
operators and spacetime coordinates. Because the spacetime coordinates 
are noncommutative in Eq. (10), we cannot apply the anticommutation 
relations for the creation and annihilation operators of Eq. (7) directly 
to obtain a $c$-number result for the Moyal anticommutator.

  In order to obtain a $c$-number result for the Moyal anticommutator, 
we can calculate its vacuum expectation value. We have 
\begin{eqnarray*}
   & ~ & \langle0\vert\{\psi_{\alpha}(x),\overline{\psi}_{\beta}
             (x^{\prime})\}_{\star}\vert0\rangle              \\
          & = & \langle0\vert\int\frac{d^{3}pd^{3}p^{\prime}}
                {(2\pi)^{3}}\frac{m}{\sqrt{E_{p}E_{p^{\prime}}}}
                \sum\limits_{s,s^{\prime}}\Big[\Big\{b(p,s)
                 u_{\alpha}(p,s)e^{-ip\cdot x},
                 b^{\dagger}(p^{\prime},s^{\prime})\overline{u}_{\beta}
                (p^{\prime},s^{\prime})
                 e^{ip^{\prime}\cdot x^{\prime}}\Big\}_\star   \\
          & ~ &  ~~~~~~~~~~~~~~~~~~   
                +\Big\{d^{\dagger}(p,s)v_{\alpha}(p,s)e^{ip\cdot x}, 
                 d(p^{\prime},s^{\prime})
                 \overline{v}_{\beta}(p^{\prime},s^{\prime})e^{-ip^{\prime}
                 \cdot x^{\prime}}\Big\}_\star\Big]\vert0\rangle     \\
          & = & \int\frac{d^{3}p}{(2\pi)^{3}}\frac{m}{E_{p}}
                \sum\limits_{s=1,2}[u_{\alpha}(p,s)\overline{u}_{\beta}(p,s)
                 e^{-ip\cdot x}\star e^{ip\cdot x^{\prime}}+
                 v_{\alpha}(p,s)\overline{v}_{\beta}(p,s)
                 e^{-ip\cdot x^{\prime}}\star e^{ip\cdot x}]         \\
          & = & \int\frac{d^{3}p}{(2\pi)^{3}2E_{p}}
                [(\not{p}+m)_{\alpha\beta}
                 e^{-ip\cdot x}\star e^{ip\cdot x^{\prime}}
               +(\not{p}-m)_{\alpha\beta}
                 e^{-ip\cdot x^{\prime}}\star e^{ip\cdot x}]       \\
          & = & \int\frac{d^{3}p}{(2\pi)^{3}2E_{p}}
                \left[(\not{p}+m)_{\alpha\beta}
                \exp(\frac{i}{2}p\times p)e^{-ip\cdot(x-x^{\prime})}
              +(\not{p}-m)_{\alpha\beta}\exp(\frac{i}{2}p\times p)
                e^{ip\cdot(x-x^{\prime})}\right] ~,
\end{eqnarray*}     
$$  \eqno{(11)}  $$   
where $p\times q=p_{\mu}\theta^{\mu\nu}q_{\nu}$ and in the last 
equality we have applied the formula (3). Because $\theta^{\mu\nu}$ 
is antisymmetric, $p\times p=0$, we obtain 
\begin{eqnarray*}
   & ~ & \langle0\vert\{\psi_{\alpha}(x),\overline{\psi}_{\beta}
             (x^{\prime})\}_{\star}\vert0\rangle              \\
   & = & \int\frac{d^{3}p}{(2\pi)^{3}2E_{p}}\left
            [(\not{p}+m)_{\alpha\beta}
            e^{-ip\cdot(x-x^{\prime})}+(\not{p}-m)_{\alpha\beta}
            e^{ip\cdot(x-x^{\prime})}\right]               \\
   & = & (i\not{\partial}_{x}+m)_{\alpha\beta}~i\Delta(x-x^{\prime})  \\
   & = & -iS_{\alpha\beta}(x-x^{\prime}) ~,
\end{eqnarray*}      
$$  \eqno{(12)}  $$  
where the singular function $\Delta(x-x^{\prime})$ is defined to be [8,9]
$$
  \Delta(x-y)= -\frac{1}{(2\pi)^{3}}\int\frac
                    {d^{3}k}{\omega_{k}}e^{i{\bf k}\cdot({\bf x}-
       {\bf y})}\sin\omega_{k}(x_{0}-y_{0}) ~. 
  \eqno{(13)}  $$
So the result of Eq. (12) is equal to the anticommutator of Dirac field 
in ordinary commutative spacetime: 
$$
  \{\psi_{\alpha}(x),\overline{\psi}_{\beta}(x^{\prime})\}
          =(i\not{\partial}_{x}+m)_{\alpha\beta}~i\Delta(x-x^{\prime}) 
          =-iS_{\alpha\beta}(x-x^{\prime}) ~.            
  \eqno{(14)}  $$
It is obvious to see that this equality relies on the antisymmetry of 
$\theta^{\mu\nu}$. We can also obtain 
$$
  \langle0\vert\{\psi_{\alpha}(x),\psi_{\beta}
             (x^{\prime})\}_{\star}\vert0\rangle = 
  \langle0\vert\{\overline{\psi}_{\alpha}(x),\overline{\psi}_{\beta}
             (x^{\prime})\}_{\star}\vert0\rangle =0 ~.          
  \eqno{(15)}  $$
When $(x-y)$ is a spacelike interval the singular function 
$\Delta(x-x^{\prime})$ is zero. Therefore we can obtain the 
vacuum expectation value for the equal-time anticommutator 
to be [9]
$$
  \langle0\vert\{\psi_{\alpha}({\bf x},t),\overline{\psi}_{\beta}
             ({\bf x}^{\prime},t)\}_{\star}\vert0\rangle =             
  -\gamma^{0}_{\alpha\beta}\partial_{0}\Delta({\bf x}-{\bf x}^{\prime},
        x^{0}-x^{\prime ~0})\vert_{x^{0}=x^{\prime ~0}}=
          \gamma^{0}_{\alpha\beta}\delta^{3}({\bf x}-{\bf x}^{\prime}) ~. 
  \eqno{(16)}  $$

  We can also calculate the expectation values between non-vacuum states 
for the Moyal anticommutator (9). Let $\vert\Psi\rangle$ represent a 
normalized non-vacuum physical state for the system of Dirac field quanta  
in the occupation eigenstate:
$$
  \vert\Psi\rangle=\vert N_{p_{1}}(s,s^{\prime})N_{p_{2}}(s,s^{\prime})
         \cdots N_{p_{i}}(s,s^{\prime})\cdots,0\rangle ~.
  \eqno{(17)}  $$ 
In Eq. (17) we use $N_{p_{i}}$ to represent the occupation number for the 
momentum $p_{i}$, and use $(s,s^{\prime})$ to represent four kinds of 
the spinors $u(p,s)$ and $v(p,s)$. $N_{p_{i}}(s,s^{\prime})$ can only take 
the values $0$ and $1$. We suppose that the occupation numbers are nonzero 
only on some separate momentums $p_{i}$. For all other momentums, the 
occupation numbers are zero. We use $0$ to represent that the occupation 
numbers are zero on all the other momentums and spins in Eq. (17). The 
state vector $\vert\Psi\rangle$ has the following properties [10]: 
$$  
  \langle N_{p_{1}}(s,s^{\prime})N_{p_{2}}(s,s^{\prime})
         \cdots N_{p_{i}}(s,s^{\prime})\cdots\vert
          N_{p_{1}}(s,s^{\prime})N_{p_{2}}(s,s^{\prime})
         \cdots N_{p_{i}}(s,s^{\prime})\cdots\rangle =1 ~,        $$
$$  
  \sum\limits_
      {{\small N_{p_{1}}(s,s^{\prime})N_{p_{2}}(s,s^{\prime})\cdots}}
      \vert N_{p_{1}}(s,s^{\prime})N_{p_{2}}(s,s^{\prime})
         \cdots N_{p_{i}}(s,s^{\prime})\cdots\rangle\langle
     N_{p_{1}}(s,s^{\prime})N_{p_{2}}(s,s^{\prime})
         \cdots N_{p_{i}}(s,s^{\prime})\cdots\vert=1 ~,             $$
$$  \eqno{(18)}  $$   
$$
  a_{s,s^{\prime}}(p_{i})\vert N_{p_{1}}(s,s^{\prime})N_{p_{2}}
       (s,s^{\prime})\cdots N_{p_{i-1}}(s,s^{\prime})0_{p_{i}}
       (s,s^{\prime})N_{p_{i+1}}(s,s^{\prime})\cdots\rangle =0 ~,      $$ 
\begin{eqnarray*}
  & ~ & a_{s,s^{\prime}}(p_{i})\vert N_{p_{1}}(s,s^{\prime})N_{p_{2}}
       (s,s^{\prime})\cdots N_{p_{i-1}}(s,s^{\prime})1_{p_{i}}
       (s,s^{\prime})N_{p_{i+1}}(s,s^{\prime})\cdots\rangle      \\ 
  & = &  (-1)^{\sum\limits_{l=1}^{i-1}N_{l}(s,s^{\prime})}     
         \vert N_{p_{1}}(s,s^{\prime})N_{p_{2}}(s,s^{\prime})
         \cdots N_{p_{i-1}}(s,s^{\prime})0_{p_{i}}(s,s^{\prime})
         N_{p_{i+1}}(s,s^{\prime})\cdots\rangle ~,
\end{eqnarray*}  
$$
  a^{\dagger}_{s,s^{\prime}}(p_{i})\vert N_{p_{1}}(s,s^{\prime})
         N_{p_{2}}(s,s^{\prime})
         \cdots N_{p_{i-1}}(s,s^{\prime})1_{p_{i}}(s,s^{\prime})
         N_{p_{i+1}}(s,s^{\prime})\cdots\rangle =0 ~,              $$
\begin{eqnarray*}
  & ~ & a^{\dagger}_{s,s^{\prime}}(p_{i})\vert N_{p_{1}}(s,s^{\prime})
         N_{p_{2}}(s,s^{\prime})
         \cdots N_{p_{i-1}}(s,s^{\prime})0_{p_{i}}(s,s^{\prime})
         N_{p_{i+1}}(s,s^{\prime})\cdots\rangle      \\ 
  & = &  (-1)^{\sum\limits_{l=1}^{i-1}N_{l}(s,s^{\prime})}     
         \vert N_{p_{1}}(s,s^{\prime})N_{p_{2}}(s,s^{\prime})
         \cdots N_{p_{i-1}}(s,s^{\prime})1_{p_{i}}(s,s^{\prime})
         N_{p_{i+1}}(s,s^{\prime})\cdots\rangle ~.
\end{eqnarray*}  
$$  \eqno{(19)}  $$ 
In Eq. (19), we use $a_{s,s^{\prime}}$ to represent one kind of the  
annihilation operators $b(p,s)$ and $d(p,s)$, and use 
$a^{\dagger}_{s,s^{\prime}}$ to represent one kind of the creation 
operators $b^{\dagger}(p,s)$ and $d^{\dagger}(p,s)$.

  From Eq. (10) the expectation value between any non-vacuum state 
$\vert\Psi\rangle$ for the Moyal anticommutator is given by   
\begin{eqnarray*}
   & ~ & \langle\Psi\vert\{\psi_{\alpha}(x),\overline{\psi}_{\beta}
             (x^{\prime})\}_{\star}\vert\Psi\rangle              \\
          & = & \langle\Psi\vert\int\frac{d^{3}pd^{3}p^{\prime}}
                {(2\pi)^{3}}\frac{m}{\sqrt{E_{p}E_{p^{\prime}}}}
                \sum\limits_{s,s^{\prime}}\Big[\Big\{b(p,s)u_{\alpha}
                (p,s)e^{-ip\cdot x},
                 b^{\dagger}(p^{\prime},s^{\prime})\overline{u}_{\beta}
                (p^{\prime},s^{\prime})
                 e^{ip^{\prime}\cdot x^{\prime}}\Big\}_\star   \\
          & ~ &  ~~~~~~~~~~~~~~~~    
                +\Big\{d^{\dagger}(p,s)v_{\alpha}(p,s)e^{ip\cdot x}, 
                 d(p^{\prime},s^{\prime})
                 \overline{v}_{\beta}(p^{\prime},s^{\prime})
                  e^{-ip^{\prime}
                 \cdot x^{\prime}}\Big\}_\star\Big]\vert\Psi\rangle     \\
          & = & \int\frac{d^{3}p}{(2\pi)^{3}}\frac{m}{E_{p}}
                \sum\limits_{s=1,2}[u_{\alpha}(p,s)\overline{u}_{\beta}
                 (p,s)e^{-ip\cdot x}\star e^{ip\cdot x^{\prime}}+
                 v_{\alpha}(p,s)\overline{v}_{\beta}(p,s)
                 e^{-ip\cdot x^{\prime}}\star e^{ip\cdot x}] ~.       
\end{eqnarray*}  
$$  \eqno{(20)}  $$
To consider the different cases of the occupation numbers for the Dirac 
field quanta, in the second equality of Eq. (20), we need to consider the 
different orders for the Moyal star-products in fact. However because 
$e^{-ip\cdot x}\star e^{ip\cdot x^{\prime}}=e^{ip\cdot x^{\prime}}
 \star e^{-ip\cdot x}$ and $e^{-ip\cdot x^{\prime}}\star e^{ip\cdot x}= 
 e^{ip\cdot x}\star e^{-ip\cdot x^{\prime}}$, we can neglect these 
difference. Therefore we obtain 
\begin{eqnarray*}
   & ~ & \langle\Psi\vert\{\psi_{\alpha}(x),\overline{\psi}_{\beta}
             (x^{\prime})\}_{\star}\vert\Psi\rangle              \\
   & = & \int\frac{d^{3}p}{(2\pi)^{3}2E_{p}}
                 [(\not{p}+m)_{\alpha\beta}
                 e^{-ip\cdot x}\star e^{ip\cdot x^{\prime}}+
                 (\not{p}-m)_{\alpha\beta}
                 e^{-ip\cdot x^{\prime}}\star e^{ip\cdot x}]      \\
          & = & \int\frac{d^{3}p}{(2\pi)^{3}2E_{p}}
                \left[(\not{p}+m)_{\alpha\beta}
                e^{-ip\cdot(x-x^{\prime})}+(\not{p}-m)_{\alpha\beta}
                e^{ip\cdot(x-x^{\prime})}\right]         \\ 
          & = & -iS_{\alpha\beta}(x-x^{\prime}) ~.
\end{eqnarray*}  
$$  \eqno{(21)}  $$
We can also obtain 
$$
  \langle\Psi\vert\{\psi_{\alpha}(x),\psi_{\beta}
             (x^{\prime})\}_{\star}\vert\Psi\rangle = 
  \langle\Psi\vert\{\overline{\psi}_{\alpha}(x),\overline{\psi}_{\beta}
             (x^{\prime})\}_{\star}\vert\Psi\rangle =0 ~,            
  \eqno{(22)}  $$
$$
  \langle\Psi\vert\{\psi_{\alpha}({\bf x},t),\overline{\psi}_{\beta}
             ({\bf x}^{\prime},t)\}_{\star}\vert\Psi\rangle =
          \gamma^{0}_{\alpha\beta}\delta^{3}({\bf x}-{\bf x}^{\prime}) ~.  
  \eqno{(23)}  $$
Thus the results of Eqs. (21) to (23) are equal to that of Eqs. (12), 
(15), and (16) of the vacuum state expectation values. We can see that 
the properties of the anticommutation relations for the creation and 
annihilation operators of Eq. (7) are still reflected in the above 
evaluations for the non-vacuum state expectation values of the Moyal 
anticommutators. Although the Lorentz invariant singular function 
$S(x-x^{\prime})$ is zero for a spacelike interval of the two spacetime 
coordinates, we cannot deduce that Dirac fields on noncommutative 
spacetime satisfy the microscopic causality principle from Eqs. (12) 
and (21) as that for the scalar field case [7], for the reason that the 
physical observables for Dirac fields are not $\psi(x)$ and 
$\overline{\psi}(x)$ directly, they are some bilinear forms constructed 
from $\psi(x)$ and $\overline{\psi}(x)$. For the microscopic causality 
problem of Dirac fields on noncommutative spacetime, we will discuss 
it in a following paper.

  The same as quantum fields on ordinary commutative spacetime [8,9], 
we can decompose the Fourier expansion of the free Dirac field into 
positive frequency part and negative frequency part: 
$$
  \psi(x)=\psi^{+}(x)+\psi^{-}(x) ~,                
  \eqno{(24)}  $$
where
$$
  \psi^{+}(x)=\int\frac{d^{3}p}{(2\pi)^{3/2}}\sqrt{\frac{m}{E_{p}}}
      \sum\limits_{s=1,2}b(p,s)u(p,s)e^{-ip\cdot x} ~,              $$
$$
  \psi^{-}(x)=\int\frac{d^{3}p}{(2\pi)^{3/2}}\sqrt
      {\frac{m}{E_{p}}}\sum\limits_{s=1,2}d^{\dagger}(p,s)v(p,s)
        e^{ip\cdot x} ~.      
  \eqno{(25)}  $$
For the conjugate field we have 
$$
  \overline\psi(x)=\overline\psi^{+}(x)+\overline\psi^{-}(x) ~, 
  \eqno{(26)}  $$
$$
  \overline\psi^{+}(x)=\int\frac{d^{3}p}{(2\pi)^{3/2}}\sqrt
      {\frac{m}{E_{p}}}\sum\limits_{s=1,2}d(p,s)\overline{v}(p,s)
       e^{-ip\cdot x} ~,          $$
$$
  \overline\psi^{-}(x)=\int\frac{d^{3}p}{(2\pi)^{3/2}}\sqrt
      {\frac{m}{E_{p}}}\sum\limits_{s=1,2}b^{\dagger}(p,s)
       \overline{u}(p,s)e^{ip\cdot x} ~.         
  \eqno{(27)}  $$
According to Eq. (11), we can decompose the vacuum expectation value 
of the Moyal anticommutator into two parts: 
$$
  \langle0\vert\{\psi_{\alpha}(x),\overline{\psi}_{\beta}
             (x^{\prime})\}_{\star}\vert0\rangle =
  \langle0\vert\{\psi^{+}_{\alpha}(x),\overline{\psi}^{-}_{\beta}
             (x^{\prime})\}_{\star}\vert0\rangle + 
  \langle0\vert\{\psi^{-}_{\alpha}(x),\overline{\psi}^{+}_{\beta}
             (x^{\prime})\}_{\star}\vert0\rangle ~,    
  \eqno{(28)}  $$
where
\begin{eqnarray*}
   & ~ & \langle0\vert\{\psi^{+}_{\alpha}(x),\overline{\psi}^{-}_{\beta}
             (x^{\prime})\}_{\star}\vert0\rangle        \\
   & = & \langle0\vert\int\frac{d^{3}p}{(2\pi)^{3}}\frac{m}{E_{p}}
                \sum\limits_{s=1,2}\Big\{b(p,s)u_{\alpha}(p,s)e^{-ip\cdot x},
                 b^{\dagger}(p,s)\overline{u}_{\beta}(p,s)
                 e^{ip\cdot x^{\prime}}\Big\}_\star \vert0\rangle  \\
   & = & \int\frac{d^{3}p}{(2\pi)^{3}}\frac{m}{E_{p}}
                \sum\limits_{s=1,2}u_{\alpha}(p,s)\overline{u}_{\beta}(p,s)
                 e^{-ip\cdot x}\star e^{ip\cdot x^{\prime}}    \\
   & = & \int\frac{d^{3}p}{(2\pi)^{3}2E_{p}}(\not{p}+m)_{\alpha\beta}
                e^{-ip\cdot(x-x^{\prime})}        \\
   & = & (i\not{\partial}_{x}+m)_{\alpha\beta}~i\Delta^{+}(x-x^{\prime}) 
         =-iS_{\alpha\beta}^{+}(x-x^{\prime}) ~,
\end{eqnarray*} 
$$  \eqno{(29)}  $$
and 
\begin{eqnarray*}
   & ~ & \langle0\vert\{\psi^{-}_{\alpha}(x),\overline{\psi}^{+}_{\beta}
             (x^{\prime})\}_{\star}\vert0\rangle        \\
   & = & \langle0\vert\int\frac{d^{3}p}{(2\pi)^{3}}\frac{m}{E_{p}}
                \sum\limits_{s=1,2}\Big\{d^{\dagger}(p,s)
                v_{\alpha}(p,s)e^{ip\cdot x},
                d(p,s)\overline{v}_{\beta}(p,s)
                 e^{-ip\cdot x^{\prime}}\Big\}_\star \vert0\rangle  \\
   & = & \int\frac{d^{3}p}{(2\pi)^{3}}\frac{m}{E_{p}}
                \sum\limits_{s=1,2}v_{\alpha}(p,s)
                \overline{v}_{\beta}(p,s)
                 e^{-ip\cdot x^{\prime}}\star e^{ip\cdot x}     \\
   & = & \int\frac{d^{3}p}{(2\pi)^{3}2E_{p}}(\not{p}-m)_{\alpha\beta}
                e^{ip\cdot(x-x^{\prime})}        
               =(i\not{\partial}_{x}+m)_{\alpha\beta}~i\Delta^{-}
                (x-x^{\prime})  \\
   & = & -iS_{\alpha\beta}^{-}(x-x^{\prime})
         =iS_{\alpha\beta}^{+}(x^{\prime}-x) ~, 
\end{eqnarray*} 
$$  \eqno{(30)}  $$
because $\Delta^{-}(x-x^{\prime})=-\Delta^{+}(x^{\prime}-x)$. 
Thus we have 
$$
  \langle0\vert\{\psi_{\alpha}(x),\overline{\psi}_{\beta}
             (x^{\prime})\}_{\star}\vert0\rangle
   = -i[S^{+}_{\alpha\beta}(x-x^{\prime})+
          S^{-}_{\alpha\beta}(x-x^{\prime})] ~,
  \eqno{(31)}  $$  
and we have defined  
$$
  S_{\alpha\beta}(x-x^{\prime})=S^{+}_{\alpha\beta}(x-x^{\prime})
           +S^{-}_{\alpha\beta}(x-x^{\prime}) ~.
  \eqno{(32)}  $$  
We can also see that if we replace the vacuum state by any 
non-vacuum state in the above formulas the results will not change. 
The above results are also equal to the corresponding anticummutators 
of Dirac fields in ordinary commutative spacetime.

  For the anticommutators of Eqs. (29) and (30), we can rewrite them 
furthermore:
\begin{eqnarray*}
   & ~ & \langle0\vert\{\psi^{+}_{\alpha}(x),\overline{\psi}^{-}_{\beta}
             (x^{\prime})\}_{\star}\vert0\rangle = 
             \langle0\vert\psi^{+}_{\alpha}(x)
       \star\overline{\psi}^{-}_{\beta}(x^{\prime})\vert0\rangle        \\
   & = & \langle0\vert\psi_{\alpha}(x)
       \star\overline{\psi}_{\beta}(x^{\prime})\vert0\rangle  
          = -iS_{\alpha\beta}^{+}(x-x^{\prime}) ~,
\end{eqnarray*} 
$$  \eqno{(33)}  $$ 
\begin{eqnarray*}
   & ~ & \langle0\vert\{\psi^{-}_{\alpha}(x),\overline{\psi}^{+}_{\beta}
             (x^{\prime})\}_{\star}\vert0\rangle = 
       \langle0\vert\overline{\psi}^{+}_{\beta}(x^{\prime})
       \star\psi^{-}_{\alpha}(x)\vert0\rangle             \\
   & = & \langle0\vert\overline{\psi}_{\beta}(x^{\prime})
       \star\psi_{\alpha}(x)\vert0\rangle         
          = -iS_{\alpha\beta}^{-}(x-x^{\prime}) ~.
\end{eqnarray*} 
$$  \eqno{(34)}  $$ 
We define the time-ordered Moyal star-product of two Dirac field 
operators to be 
$$
  T\psi_{\alpha}(x)\star\overline{\psi}_{\beta}(x^{\prime})=
   \theta(t-t^{\prime})\psi_{\alpha}(x)\star
   \overline{\psi}_{\beta}(x^{\prime})
   -\theta(t^{\prime}-t)\overline{\psi}_{\beta}
   (x^{\prime})\star\psi_{\alpha}(x) ~,         
  \eqno{(35)}  $$ 
where $\theta(t-t^{\prime})$ is the unit step function. We can 
calculate the vacuum expectation value of Eq. (35): 
$$
  \langle0\vert T\psi_{\alpha}(x)\star\overline{\psi}_{\beta}
        (x^{\prime})\vert0\rangle =
   \theta(t-t^{\prime})\langle0\vert\psi_{\alpha}(x)
    \star\overline{\psi}_{\beta}(x^{\prime})\vert0\rangle 
   -\theta(t^{\prime}-t)\langle0\vert\overline{\psi}_{\beta}
   (x^{\prime})\star\psi_{\alpha}(x)\vert0\rangle ~.          
  \eqno{(36)}  $$
Equation (36) is just the Feynman propagator of Moyal star-product 
for Dirac field. We can call it the Feynman Moyal propagator for 
convenience. To introduce the singular function $S_{F}(x)$, 
we can write the Feynman Moyal propagator (36) as  
$$
  \langle0\vert T\psi_{\alpha}(x)\star\overline{\psi}_{\beta}
        (x^{\prime})\vert0\rangle=iS_{F}(x-x^{\prime})_{\alpha\beta} ~. 
  \eqno{(37)}  $$
From Eqs. (33), (34), and (36), we have 
$$
  S_{F}(x-x^{\prime})_{\alpha\beta}=-[\theta(t-t^{\prime})
       S^{+}_{\alpha\beta}(x-x^{\prime})
       -\theta(t^{\prime}-t)S^{-}_{\alpha\beta}(x-x^{\prime})] ~,
  \eqno{(38)}  $$
where the momentum integral representation for the singular function 
$S_{F}(x-x^{\prime})$ is given by 
$$
  S_{F}(x-x^{\prime})=\int\frac{d^{4}p}{(2\pi)^{4}}
       \frac{\not{p}+m}{p^{2}-m^{2}+i\epsilon}
        e^{-ip\cdot(x-x^{\prime})} ~.  
  \eqno{(39)}  $$
From the above results, we can see that the Feynman Moyal propagator 
of Dirac field on noncommutative spacetime is just equal to the 
Feynman propagator of Dirac field on ordinary commutative spacetime. 
However it is necessary to point out that in Eq. (35) for the 
definition of the time-ordered Moyal star-product of two Dirac fields, 
we have made a simplified manipulation. This is because the Moyal 
star-products are not invariant generally under the exchange of the 
orders of two functions, for the second term of the right hand side 
of Eq. (35) we need to consider this fact. In the Fourier integral 
representation, it can be seen clearly that the second term of the 
right hand side of Eq. (35) will have an additional phase factor 
$e^{ip\times p^{\prime}}$ in contrast to the first term of the right 
hand side of Eq. (35) due to the exchange of the order of 
$\psi_{\alpha}(x)$ and $\overline{\psi}_{\beta}(x^{\prime})$ for the 
Moyal star-product. However in Eq. (36) when we calculate the vacuum 
expectation value for Eq. (35), we can see that the non-zero 
contribution comes from the $p=p^{\prime}$ part inside the integral 
(cf. Eqs. (29) and (30)). This will make the phase factor to be 
$e^{ip\times p}$, which is $1$ due to the antisymmetry of 
$\theta^{\mu\nu}$. Thus in the right hand side of Eq. (35), we can 
omit this effect in the exchange of the order of two Dirac fields 
for their Moyal star-product equivalently.

  Just like that in ordinary commutative spacetime, the physical 
meaning of the Feynman Moyal propagator (36) can also be explained 
as the vacuum to vacuum transition amplitude for Dirac fields on 
noncommutative spacetime. The reason why we would like to construct 
the Feynman propagators of Moyal star-products for quantum fields on 
noncommutative spacetime is that: for noncommutative field theories 
we can establish their $S$-matrix where the products between field 
operators in ${\cal H}_{int}$ are Moyal star-products. From the 
Wick's theorem expansion for the time ordered products of field 
operators, there will occur the Feynman Moyal propagators. Therefore 
it is necessary to study Feynman propagators of Moyal star-products 
for quantum fields on noncommutative spacetime.

\vskip 1cm

\noindent {\large {\bf References}}

\vskip 12pt

[1] H.S. Snyder, Phys. Rev. {\bf 71}, 38 (1947). 

[2] A. Connes, {\sl Noncommutative geometry} (Academic Press, 
    New York, 1994). 

[3] S. Doplicher, K. Fredenhagen, and J.E. Roberts, Phys. Lett. B 
    {\bf 331}, 39 (1994); 

    ~~~ Commun. Math. Phys. {\bf 172}, 187 (1995), hep-th/0303037. 
    
[4] N. Seiberg and E. Witten, J. High Energy Phys. 09 (1999) 032, 
    hep-th/9908142, and 

    ~~~ references therein.   

[5] M.R. Douglas and N.A. Nekrasov, Rev. Mod. Phys. {\bf 73}, 977 
    (2001), hep-th/0106048. 
 
[6] R.J. Szabo, Phys. Rep. {\bf 378}, 207 (2003), hep-th/0109162. 

[7] Z.Z. Ma, hep-th/0601046. 

[8] J.D. Bjorken and S.D. Drell, {\sl Relativistic Quantum Fields}  
    (McGraw-Hill, 1965). 

[9] C. Itzykson and J.-B. Zuber, {\sl Quantum Field Theory} 
    (McGraw-Hill Inc., 1980). 

[10] L.D. Landau, E.M. Lifshitz, Quantum Mechanics, 
     Pergamon Press, 1977.

\end{document}